Chapter 1

# INTERNET OF THINGS FORENSICS: CHALLENGES AND CASE STUDY

Saad Alabdulsalam, Kevin Schaefer, Tahar Kechadi and Nhien-An Le-Khac

**Abstract**   Today is the era of Internet of Things (IoT), millions of machines such as cars, smoke detectors, watches, glasses, webcams, etc. are being connected to the Internet. The number of machines that possess the ability of remote access to monitor and collect data is continuously increasing. This development makes, on one hand, the human life more comfortable, convenient, but it also raises on other hand issues on security and privacy. However, this development also raises challenges for the digital investigator when IoT devices involve in criminal scenes. Indeed, current research in the literature focuses on security and privacy for IoT environments rather than methods or techniques of forensic acquisition and analysis for IoT devices. Therefore, in this paper, we discuss firstly different aspects related to IoT forensics and then focus on the current challenges. We also describe forensic approaches for a IoT device - smartwatch as a case study. We analyze forensic artifacts retrieved from smartwatch devices and discuss on evidence found aligned with challenges in IoT forensics.

## 1.     Introduction

Internet of things (IoT) is a new revolution of technology that enables small devices to act as smart objects. These devices are connected with each other by different network media types, and the result of these communications is to return to the sensors by appropriate decision. The goal of IoT is to make lives more convenient and dynamic. For instance, cars can drive alone, the smart light turned off when there is no one in the room, air conditioner turned on when room temperature goes be-low a certain degree. Moreover, IoT devices can exchange information between themselves to provide a convenient services to the owner. For instance, a smart player can select and play a particular song dependent on the blood pressure of owner which was taken from his/her smartwatch. Indeed, IoT technology crosses different industry areas such as smart city, medical care, social domains and smart home [1].

However, the IoT technology can create more opportunities for cybercrimes to attack these areas, resulting in a direct impact on users. In addition, as most of all consumer technology, IoT technology is not designed with security in mind, as the main concern was to minimize the cost and size. Therefore, these devices have shortage in hardware resources. Because of this lack, most of the security tools cannot be installed in IoT devices, as they require a certain space and process function to be run [2], making them an easy target for cybercrimes. They can find the way to inject these devices in order to use them as weapons to attack other websites [3]. Cybercrimes with the power of IoT technology can cross the virtual space to threaten human life and with increasing number of these crimes, which consider as one of the two main reasons why we need IoT forensic. For instance, in January last year, FDA warned that certain pacemakers ,which is a system that sends electrical impulses to the heart in order to set the heart rhythm, are vulnerable to hacking[4]. This means who used the vulnerable device in that particular time could face death, if his/her life became under a hacker who could control the pacemaker. The second main reason is IoT digital evidence is a rich and often unexplored source of information. As most of IoT manufacturers show to the customers what the product can provide to them but they do not mention about the process that happened to provide these services. For example: a LG smart vacuum, which can clean the room by itself, it seems that the function is done its job by sensors which can detect the size and the shape of the roam then start clean it. However, a group of researcher found a vulnerability in an LG portal login process that allowed them to take control of the vacuum and consequently, giving them access to live-stream video from inside a home [5]. This incidence rises some important questions, does the LG portal record cleaning process at every time the vacuum is running? And where is stored! also where the process is occurred! Is locally or through the Cloud?

From the forensic perspective, each IoT device will provide important artifacts that could help in the investigation process. Some of these artifacts have not been disclosed in public that means the investigators should consider of these resources and how they can acquire the artifacts from these devices. Even though IoT has rich sources of evidence from the real world application, it causes some challenges for forensics examiner including but not limited to the location of data and heterogeneous nature of IoT devices such as differences in operating systems and communication standards [6]. Current research in literature focuses on IoT security and privacy, however, some important aspects such as Incident response and forensic investigations, have not been covered efficiently. Therefore, this paper will spot the light on these aspects. In this paper, we discuss on IoT forensics and how is different from traditional forensics and following with what are the challenges of IoT forensics. We also describe the forensic acquisition of smartwatch, an IoT device as a case study. The rest of paper is organized as follows: Section 2 shows the difference between the traditional forensics vs. IoT forensics. We then discuss on IoT forensics challenges in Section 3. We present our forensics acquisition and analysis of Apple Smart Watch in Section 4. Finally, we conclude and show future work in Section 5.

## 2. Traditional digital forensics vs. IoT Forensics

Digital forensics could be defined as a processing that use to identify the digital evidence in its most original form and then performing a structured investigation to collect, examine and analyze the digital evidence. There are several aspects of difference and similarity between traditional and IoT forensics. In terms of evidence sources, traditional evidence could be computers, mobile devices, servers or gateways. In IoT forensics, the evidence could be home appliances, cars, tags readers, sensor nodes, medical implants in humans or animals, or other IoT devices.

In terms of Jurisdiction and Ownership, there are no differences, it could be individuals, groups, companies, governments, etc. In terms of evidences data types, IoT type could be any possible format, it could be a specific format for a particular vendor. However, in traditional forensics could be electronic documents or standard file formats. In terms of networks, IoT limitation encourages to be seen a new protocol that fits this limitation. However, the network boundaries become not clear as the traditional network. Increasing in the blurry boundary lines, it makes seizing IoT forensic devices become one of the challenges of IoT forensics. In [7], authors highlight this issue and introduced possible identification for the source of evidence.

## 3. IoT Forensics

IoT technology is a combination of many technology zones: IoT zone, Network zone and Cloud zone. These zones can be the source of IoT

Digital evidences. That is, an evidence can be collected from a smart IoT device or a sensor, from an internal network such as a firewall or a router, or from outside networks such as Cloud or an application. Based on these zones, IoT has three aspects in term of forensics: Cloud forensics, network forensics and device level.

Most of IoT devices have the ability to cross Internet (direct or indirect connect) through applications to share their resources in the Cloud. With all valuable data that store in the Cloud, it has recently became one of the most important targets for attackers. In traditional digital forensics, the examiner can hold the digital equipment and then apply the investigation process to extract the evidence. However, in Cloud forensics [8], it is a different scenario, the evidence could be separated in multi-location which is rising many challenges in terms of acquisition of data from the Cloud. In addition, in the Cloud, examiners have limited control and access to seize the digital equipment and getting an exact place of evidence could be a challenge [9]. Dykstra also addressed this challenge in one of the case study that provided about child pornography website. In the warrant that request the Cloud provider, should provide the name of the data owner, or specify the location of the data that you are looking for [10]. Besides, data could be stored in a different location in the Cloud, resulting in no evidence could be seized. In addition, as all Cloud services use Virtual Machine as servers, data volatile like registry entries or temporary Internet files in these servers could be erased if they not synchronized with storage devices. For instance, if these servers are restarted or shutdown, the data could be erased.

Network Forensics include all different kinds of networks that IoT devices used to send and receive data. It could be home networks, industrial networks, LANs, MANs and WANs. For instance, if an incident occurs in IoT devices, all logs that traffic flow that has passed throw, could be potential evidence such as firewalls or IDS logs [11].

Device Level Forensics include all potential digital evidence that can be collected from IoT devices like graphics, audio, video [12][13]. Videos and graphics from CCTV camera or audios from Amazon Echo, can be great examples of digital evidences in the device level forensics.

### 3.1   IoT forensic Challenges

IoT technology has presented a significant shift in investigation field, especially in how it interacted with data. However, there are some challenges in terms of IoT forensics.

**Data Location**   Many of IoT data are spread in different locations which are out of the user control. This data could be in the Cloud,

in third party's location, in mobile phone or other devices. Therefore, in IoT forensics, to identify the location of evidence is considered as one of the biggest challenges can investigator faced in order to collect the evidence. In addition, IoT data might be located in different countries and be mixed with other users information, which means different countries regulations are involved [14]. A great case example is what was happened in August 2014, when a Microsoft refused to comply with a search warrant that seeking data stored outside the country of warrant(US), making the case opened for a long period of time[15].

**Lifespan limitation of digital media**   Because the limitation of storage in IoT devices, the lifespan of data in IoT devices is short and data can be easily overwritten. Resulting in the possibility of evidence being lost [16]. Therefore, one of the challenges is the period of survival of the evidence in IoT devices before it is overwritten. Transferring the data to another thing such as local Hub or to the Cloud could be an easy solution to solve this challenge. However, it present another challenge that related to securing the chain of evidence and how to prove the evidence has not been changed or modified [16].

**Cloud service requirement**   Most of the accounts are anonymous users because Cloud service does not require the accurate information from user to sign up for their service. It could lead to impossible to identify a criminal [17] . For example, even though the investigators find an evidence in the Cloud that prove a particular IoT device in crime scene is the cause of the crime, it does not mean this evidence could lead to identify the criminal.

**Security lack**   Evidence in IoT devices could be changed or deleted because of lack of security, which could make these evidence not solid enough to be accepted in law court[18][19] .For example, in the market, some companies do not update their devices regularly or at all or sometime they stop supporting the device's framework when they focus on a new product with the new infrastructure. As a result, it could leave these devices vulnerable as hacker found a new vulnerability.

**Device type**   In identification phase of forensics, the digital investigator needs to identify and acquire the evidence from a digital crime scene. Usually, evidence source is types of a computer system such as computer and mobile phone. However, in IoT, the source of evidence could be objects like a smart refrigerator or smart coffee maker. [7]. Therefore, the investigators will face some challenges. One of these challenging is Iden-tifying and finding the IoT devices in crime scene. It could the device terned off because it run out of battery, which make the chance to be found is so difficult especially if the IoT devices is very small, in hidden place or look likes a traditional device. Carrying the device to the lab and finding a space could be anther challenge that investigator could face in terms of device type. In addition, extracting the evidences form these devices is considered as anther IoT challenges as most of manufacturer adopts different platforms, operating systems and hardwares. One of the examples is the CCTV forensics [20] where the CCTV's manufacturers applied different file system format in their devices. Retrieving properly artifacts from CCTV's storage devices is still a challenges. We also show in [21] a new approach to carve the deleted video footprint a proprietary designed file storage system.

**Data Format** The format of the data that generated by IoT devices is not matching to what is saved in the Cloud. In addition, user have no direct access to his/her data and the data presents in deferent format than that in witch it is stored. Moreover, Data could be process using analytic functions in different places before be stored in the Could. Hence, in order to be accepted in a law court, data form should be returned to original format before performing analysis[7].

## 3.2 Limitations in the Currently Available Forensic Tools

The existing tools in digital forensics field cannot fit with the heterogeneous infrastructure of IoT environment. The massive amount of possible evidence that are generated by a large number of IoT devices, it will consequently bring new challenges in the aspect of collecting evidence from distributed IoT infrastructures. In addition, since a hacker can monopolize the evidence in IoT devices because the weakness of these devices in term of security, the extracting evidence from them maybe not acceptable in law court [9]. Moreover, because most of IoT data are stored in the Cloud, the Cloud becomes one of the main sources of evidence in IoT. Hence, investigators will face some of the problems of collecting evidence from the Cloud, because the procedures of digital forensic and tools assume to have physical access to the evidence source. However, in the Cloud, the investigators could find a difficulty to even to know where the data is located [9]. In addition, the physical servers could have many virtual machines that belong to different owners. Moreover, Cloud environments could not be available when a crime has been committed. Therefore, all of these challenges need to be addressed and to find a method to work around barriers and come up with a new tool for IoT investigation, that can be approved by law court and achieve of investigators goals [12].

## 4. A Case Study : Smartwatch Forensics

In this section, we present a case study on an IoT device: an Apple smartwatch to show that approaches for the forensic acquisition and analysis of IoT devices are still device-oriented. A smartwatch is a digital wristwatch and a wearable computing device. Digital watches are categorized according to their technologies. There are smartwatches, activity trackers, running, multisport and smart home. Because of the ongoing development of digital watches, most watches cover more than one technology. A smartwatch is used like a smartphone and has mostly similar functions. Among others, a smartwatch shows the date and time, counts steps and provides various types of information. It can provide news, weather reports, flight information, traffic news, and receive text messages, e-mails, social media messages, tweets and many others. The connectivity of smartwatches plays an important role in the retrieval of collected data and information from the Internet. To be a full-featured smartwatch, a watch must have a good connectivity to enable it to communicate with other devices (e.g. a smartphone) and be able to work independently. In this case study, we investigate Apple Watch Series 2 with the following technical specification details:

- Network-accessible smartwatch, no cellular connectivity.
- Dual-Core Apple S2 chip.
- Non-removable, built-in rechargeable lithium-ion battery.
- watchOS 2.3, watchOS 3.0, upgradable to watchOS 3.2.
- Wi-Fi 802.11 b/g/n 2.4GHz, Bluetooth 4.0, Built-in GPS, NFC chip, Service Port.
- AMOLED capacitive touchscreen, Force Touch, 272 by 340 pixels (38 mm), 312 by 390 pixels (42 mm), Sapphire crystal or Ion-X glass.
- Sensors: accelerometer, gyroscope, heart-rate sensor, ambient light sensor
- Messaging: SMS (tethered), email, iMessage
- Sound: vibration, ringtones, loudspeaker

It is important to note that the Apple Watch Series 2 has a hidden diagnostic port [22]. No official cable is available for it, so our approach is doing the forensics via the iPhone synchronized with the Apple Watch including: (i) logical acquisition with Cellebrite UFED of relevant data from an Apple iPhone; (ii) Manual swipe through the Apple Watch. Basically, the investigators are interested in the following artifacts: GPS-data, heart-rate data, timestamps, MAC address, paired devices, text messages and emails, call log, contact, etc.

## 4.1 Logical Acquisition

The following findings on iPhone in relation of Apple Watch are the result of multiple extractions to clarify attempts and changes. The first hint of an Apple Watch is given in the database
 com.apple.MobileBluetooth.ledevices.paired.db
This database can be found in the following path in the file system of the iPhone:
 /SysSharedContainerDomain- systemgroup.com.apple.bluetooth/Library/Database/
This Database contains the UUID, name, address, resolved address, LastSeenTime and LastConnectionTime. Since the Apple Watch does not have a separate file system on the iPhone, Apple Watch data has to be searched for the inside application data on the iPhone. In our case, the Apple Watch was used in the operation time with following applications: Health App, Nike Plus App, Heartbeat App, Messages, Maps App. We describe the artifacts retrieved from some of these applications in the following sub-sections.

**Health App:** The healthdb.sqlite database on Path /var/mobile/Library/Health/ contains the Apple Watch as a source device for health data (1).

**Nike Plus GPS:** The Nike Plus GPS app contains a folder com.apple.watchconnectivity. The path is /Applictions/com.nike.nikeplus-gps/Documents/inbox/. There is a folder named 71F6BCC0-56BD-4B4s-A74A-C1BA900719FB. This indicates the use of an Apple Watch. The main database in the Nike Plus GPS application is activityStore.db in Path /Applications/com.nike.nikeplus-gps/Documents/. ActivityStore.db contains an activity overview, last-ContiguosActivity, metrics, summaryMetrics and tags, which are highly relevant for the investigation.

**GPS Data:** The GPS data found in the tables metrics and tags is shown in detail below. Longitudes and latitudes are generated by the Nike Plus application and saved into the tables with a related timestamp.

Based on this information, we can create map with the GPS data in Google maps (2).

Analysis: For the logical acquisition, the software Cellebrite UFE.O and 4PC were mainly used. These tools helped to perform a logical acquisition. In the file system of the iPhone, information (i.e. UUID and name) of the paired Apple Watch could be found. Indeed, information about last connection was also found. After retrieving information about the watch, we investigated the file system in relation to the applications used with the Apple Watch. Some applications contain information about the paired Apple Watch as a source device as well. A lot of data on the iPhone were generated by the Apple Watch and come from the Apple Watch.

In summary, the iPhone contains data about workouts that have been manually started by the user while wearing the Apple Watch. But heart rate data, steps data and sleep data are recorded while wearing the Apple Watch even when no applications have been started manually. All data are provided with timestamps in different formats. In addition, the survey of law enforcement authorities and previous explorations revealed that GPS data were never found on or in relation to smartwatches. In the practical approach, GPS data generated from the application Nike+ GPS on the Apple Watch was found on the iPhone.

### 4.2 Manual Acquisition

To determine what data is stored on the Apple Watch Series 2, a manual investigation in the form of manual acquisition through device screen is used. This method of investigation was used because no physical access was possible. This variant was used to prove that the watch not only generates data but also contains data directly on the watch and can be used as an independent device. Before using the watch as an independent device, it must have been paired with an iPhone and authenticated on the same WiFi network. After this process had been carried out, the iPhone was turned off. This is only necessary to write messages or emails, or to take phone calls. For any other forensic acquisition, as is shown in the next sections, no connection to the watch is needed.

**Message**  First the messages were checked. It was possible to look at all iMessages and text messages that had been synchronized to the watch before turning off the iPhone. After putting the watch in flight mode, text messages and iMessages could still be read. We also tried to write dictated iMessages and text messages directly from the watch to any person, with the flight mode off. It was possible to send iMessage directly from the watch to the receiver. Indeed, text messages could be written on the watch. However, a click on the send button did not send the message, but rather saved it on the watch. After turning on the iPhone, this message was sent as well.

**Pictures**  Pictures were also synchronized to the watch before turning off the phone. To prove that copies of the pictures were undoubtedly on the watch and not in the Cloud, the watch was put in flight mode. The pictures were still on the watch.

**Application**  The applications HeartRate, HeartWatch, Activity, Maps, Workout, Nike+ Run, Twitter and Instagram were browsed through. The HeartRate application contains only data about the last and current heart rate measure. HeartWatch, a third party application, contains a little more data. This includes pulse, daily average, training data and sleep tracking data. The application for recording any kind of workout is Apples Workout. This application shows very little data about the last workout done and recorded. It only contains the type, length and date of the workout. Nike+ Run also contains very little data. Only the distance run in the last workout is displayed to the investigators. Both Twitter and Instagram can only be used if an iPhone is connected to the watch. If the iPhone is turned off, an indication that no phone is connected is shown.

**Email**  Reading e-mails on the watch works in the same way as reading iMessages and text messages. When the iPhone is powered off, mails can be received, opened and sent simply and independently from the iPhone. After putting the Apple Watch in flight mode, as can be seen from the plane icon, mails can still be read.

**Calendar**  In the calendar application, it is possible to see entries made by the user, but this is restricted to entries starting from the day before manual acquisition until seven days in the future. This is also possible in flight mode.

**Contacts and Phone**  Contacts are also saved on the watch independent of the status of the iPhone. The phone can be turned off and the watch can be disconnected from all networks, but contacts remain on the watch. All contacts can be displayed with all contact details saved on the iPhone. The phone application also contains a call log and the favorites list. Although the iPhone is powered off and the watch is in flight mode, the investigator has the possibility to see all voicemails and listen to them. Additionally, the phone number origin and the date and time of the incoming voicemail are displayed. After clicking the play button, the voicemail is played.

**Analysis** Since physical access to the Apple Watch is not available, manual acquisition in the form of a search through the screen is currently the only method to see what is saved and stored on the Apple Watch itself. Our research shows that the watch can be used as a standalone device independent of the iPhone. Furthermore, many artifacts that are very important for investigators, as the survey revealed, were found on the Apple Watch. These include artifact of iMessages, text messages, pictures, workout data, heart rate data, map search data, emails, calendar entries, contacts, call logs, and voicemails. For manual acquisition to be possible, a smartwatch must be unlocked. If an Apple Watch is locked with a pin code there are no options to unlock the smartwatch except using the correct pin code.

## 5. Conclusion and Future Work

In this paper we present firstly different aspects related to IoT forensics as well as challenges in acquiring and analyzing evidence from IoT devices. We also reviewed some current approaches in this context. However, most of them only focuses on the extension of traditional forensics process to adapt the IoT forensics. Through our case study, we notice that current digital forensic tools can be used at some stages in whole IoT process but there is still missing a general and efficient IoT forensic model or process that could assist the investigators to cope with these challenges today. That is also what we are working on as a future work.